\documentclass[%
aps, reprint,
showpacs,
preprintnumbers,
 amsmath,
 amssymb,
 prl,
]{revtex4-1}

\usepackage{graphicx}
\usepackage{dcolumn}
\usepackage{bm}

\usepackage{mathrsfs}
\usepackage{amssymb}
\usepackage{graphicx}

\begin{document}

\def\bm{\boldsymbol}

\def\dl{\displaystyle}
\def\du{\end{document}}
\def\d{{\rm d}}
\def\e{{\rm e}}
\def\i{{\rm i}}

\title{Two-stage acceleration of interstellar ions driven by high-energy
lepton plasma flows}

\author{CUI YunQian$^1$}{}
\author{SHENG ZhengMing$^{2,3,4}$}\email{zmsheng@sjtu.edu.cn}
\author{LU QuanMing$^5$}
\author{LI YuTong$^{1,4}$}
\author{ZHANG Jie$^{3,4}$}

\affiliation{$^1$ Beijing National Laboratory of Condensed Matter
Physics, Institute of Physics, CAS, Beijing 100190, China;}
\affiliation{$^2$ SUPA, Department of Physics, University of
Strathclyde, Glasgow G4 0NG, United Kingdom;} \affiliation{$^3$
Key Laboratory for Laser Plasmas (MoE) and Department of Physics
and Astronomy, Shanghai Jiao Tong University, Shanghai 200240,
China;} \affiliation{$^4$ IFSA Collaborative Innovation Center,
Shanghai Jiao Tong University, Shanghai 200240, China}
\affiliation{$^5$ School of Earth and Space Sciences, University
of Science and Technology of China, Hefei, 230026, China}

\date{\today}


\begin{abstract} We present the particle-in-cell (PIC) simulation
results of the interaction of a high-energy lepton plasma flow
with background electron-proton plasma and focus on the
acceleration processes of the protons. It is found that the
acceleration follows a two-stage process. In the first stage,
protons are significantly accelerated transversely (perpendicular
to the lepton flow) by the turbulent magnetic field "islands"
generated via the strong Weibel-type instabilities. The
accelerated protons shows a perfect inverse-power energy spectrum.
As the interaction continues, a shockwave structure forms and the
protons in front of the shockwave are reflected at twice of the
shock speed, resulting in a quasi-monoenergetic peak located near
200MeV under the simulation parameters. The presented scenario of
ion acceleration may be relevant to cosmic-ray generation in some
astrophysical environments.

\vspace{3mm}

\end{abstract}


\maketitle






\section{Introduction}\label{sec:intro}
The origin of high-energy cosmic-rays over a wide range of
energies and the non-thermal emission of radiation from a wide
variety of high energy astrophysical sources is the most
fundamental problem in astrophysics and has been studied for over
six decades since Teller and Fermi\cite{FermiIIOrigin}, however,
it has not been fully resolved till now. Usually these radiations
are attributed to mechanisms like inverse Compton
scattering\cite{InvCompton1, InvCompton2} and/or synchrotron
emission\cite{synchro} from high energy particles and the problem
turns to how these power-spectrum energetic particles are
generated. Considerable efforts are devoted to finding possible
generation mechanisms of such energetic
particles\cite{BhattacharjeePR2000,OlintoPR2000}. As far back as
1949, Fermi purposed the stochastic acceleration mechanism
suggesting that the particles can be accelerated through the
collision with magnetic "islands" in the
space\cite{FermiIIOrigin}. Recently, Hoshino presented a similar
process using a magnetic reconnection configuration in pair
plasmas \cite{FermiIIRecon}. Another hopeful candidate of "cosmic
particle accelerator" is the collisionless shock. It is believed
to exist widely in the interstellar space and play an important
role in supernova remnants (SNRs)\cite{ShockSNR,WangB}, jets of
radio galaxies \cite{JofR}, gamma-ray bursts
(GRBs)\cite{GRB1,GRB2} and the formation of the large scale
structure of the Universe\cite{FofUni1,FofUni2}. Therefore, it has
been modelled by plenty of researchers both analytically
\cite{ShockA1} and numerically\cite{AstroShock,ShockS2}.\par

Although both stochastic acceleration and collisionless shock
acceleration have been well modelled separately in lots of
publications, most of them are restricted to the phenomena
themselves. Presumed particular initial conditions are used to
ensure the occurrence of the interested phenomenon. For example,
in Hoshino's work \cite{FermiIIRecon}, four Harris current sheets
are imposed to build magnetic islands; in the collisionless shock
simulations\cite{AstroShock,ShockS2,ShockS3,YangJGR2009}, a
reflecting boundary condition is used to "make" the shock. These
studies have significantly improved our general understanding of
the stochastic acceleration and shock dynamics. However, more
realistic modelling in a wide parameter range is still needed.
\par
In this work, we present a scenario of proton acceleration
processes in two distinct processes occurring  naturally during
the interaction of a high-energy lepton plasma flow or jet with
background electron-proton plasma. This system of interaction
corresponds to several astrophysical models such as the
pulsar-wind shock system\cite{pulsarwind}, the fireball model of
GRBs\cite{Fireball} and the recently purposed Binary orbits model
of novae $\gamma$-ray emission\cite{BinaryOrbit}. Also the
electron-positron jets are found in different astrophysical
environments such as quasars \cite{jet1}, black holes \cite{jet2},
and so on. Without using any artificial setup to facilitate any
specific acceleration process, by use of the particle-in-cell
(PIC) simulation, we find that the background protons in the
system are naturally accelerated via Fermi acceleration like
processes. The acceleration can be divided into two stages which
are dominated by different mechanisms \cite{CuiYQ2015}. In the
early stage, stochastic acceleration, which is driven by turbulent
magnetic fields generated by strong Weibel-type instabilities,
dominates and the protons are accelerated transversely. As time
elapses, a collisionless shock gradually forms in the background
plasma, which accelerates the protons longitudinally. These
results provide a clear physics picture of a high-energy plasma
flow (jet, pulsar-wind, and so on) interaction with interstellar
masses and thus can improve our understanding of the related
astrophysical phenomena.\par

\vspace{-1mm}
\section{Simulation setup}\vspace{-1mm}
The numerical simulation was performed by our self-encoded
two-dimensional PIC code KLAPS \cite{WangPRE2015}.  We simulate a
two-dimensional unmagnetized system with the electron-ion
background plasma which homogeneously fills the whole simulation
domain located at $(x,y)\in [0,900d_e]\times [0, 95d_e]$, where
$d_e = c/\omega_{pe} = (4\pi e^2n_p/m_ec^2)^{-1/2}$ is the
election skin depth for a number density $n_p$ and electron mass
$m_e$; $e$ and $c$ here denote the elementary charge and the speed
of light in vacuum, respectively. We use a real ion-electron mass
ratio in the simulation, i.e., $m_i/m_e=1836$. The initial
temperature for background plasma is 500eV. A monoenergetic pair
plasma flow consists of electrons and positrons which are injected
along $+x$ direction from the left boundary of the simulation
domain with initial energy $\sim 50$MeV($\gamma_{e^+,e^-} = 100$)
and interact with the background plasma. The initial density of
the flow $n_f$ is equal to $n_p$. The whole system is numerically
resolved with 8 cells per $d_e$ in both directions with 9
particles per cell for each species, thus ensuring that the
behavior of leptons can be modelled accurately. The time
resolution of the simulation is $0.06\omega_{pe}^{-1}$. An open
boundary condition is used for $x$-direction and a periodic
boundary condition for $y$-direction.\par

\section{Results and discussions}
\subsection{First stage: stochastic acceleration}
\begin{figure}
\includegraphics[width=0.48\textwidth]{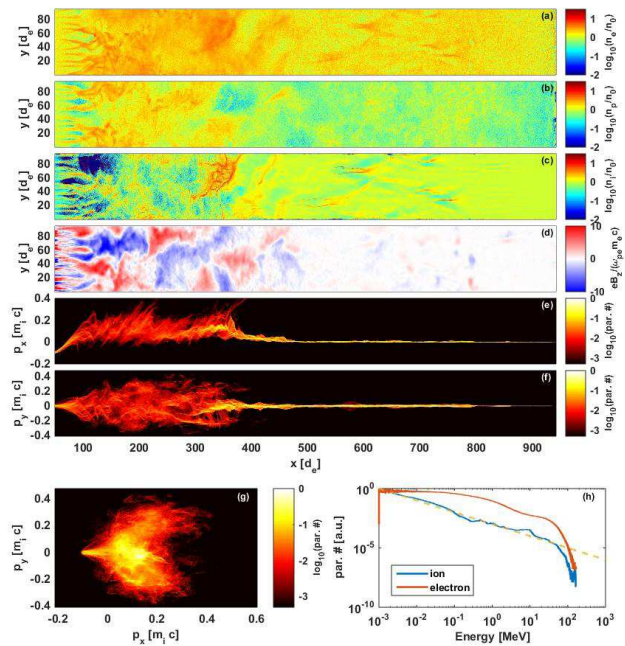}
\caption{The space distributions of (a)electrons, (b)positrons,
(c)ions and (d)magnetic fields as well as (e)$x-p_x$, (f)$x-p_y$
and (g)$p_x-p_y$ projections of the ion's phasespace distribution
at $t=950\omega_{pe}^{-1}$. (h) shows the energy spectrum of {\it
background} ions (blue line) and electrons (orange line) at the
same time, with the line $\sim E^{-1}$(dashed line).}
\label{fermi2fig}
\end{figure}

In the beginning, the flow is uniform along the $y$-direction.
However, due to the relative motion between the lepton flow and
background, Weibel-type instabilities can occur and generate
filamentous structures at $x<100d_e$ after several hundreds of
$1/\omega_{pe}$. These filaments make the newly incoming flow
leptons filamented, as shown in Figure \ref{fermi2fig}(a)-(b).
This filamented high-energy lepton plasma flow then propagates
deeper and causes a highly turbulent region at 100-400$d_e$ after
about $900\omega_{pe}^{-1}$. Within the turbulent region, the
density of ions can vary more than three orders of the magnitude,
as shown in Figure \ref{fermi2fig}(c). Such strong turbulence can
generate strong magnetic fields which are displayed in Panel (d)
of Figure \ref{fermi2fig} \cite{Mondalturb}. As said before, the
 paper mainly focuses on the acceleration of the ions in
the background plasma (referred as "ions" after for simple).
Figures \ref{fermi2fig}(e)-(g) gives the phasespace distribution
of the ions. The ions are accelerated in both $x$ and $y$
directions. The longitudinal acceleration in $x$ direction can be
simply attributed to the momentum transfer between the flow
leptons and the background particles through the electromagnetic
fields. However, it can be seen from Figure \ref{fermi2fig}(g)
that the transverse acceleration is even stronger than the
longitudinal acceleration. Moreover, the accelerated protons forms
a perfect inverse-power energy spectrum
\begin{equation}
dN/dE\sim E^{-1}
\end{equation}
with the cutoff energy near 100 MeV, as shown in Figure
\ref{fermi2fig}(h).\par

\begin{figure}
\includegraphics[width=0.48\textwidth]{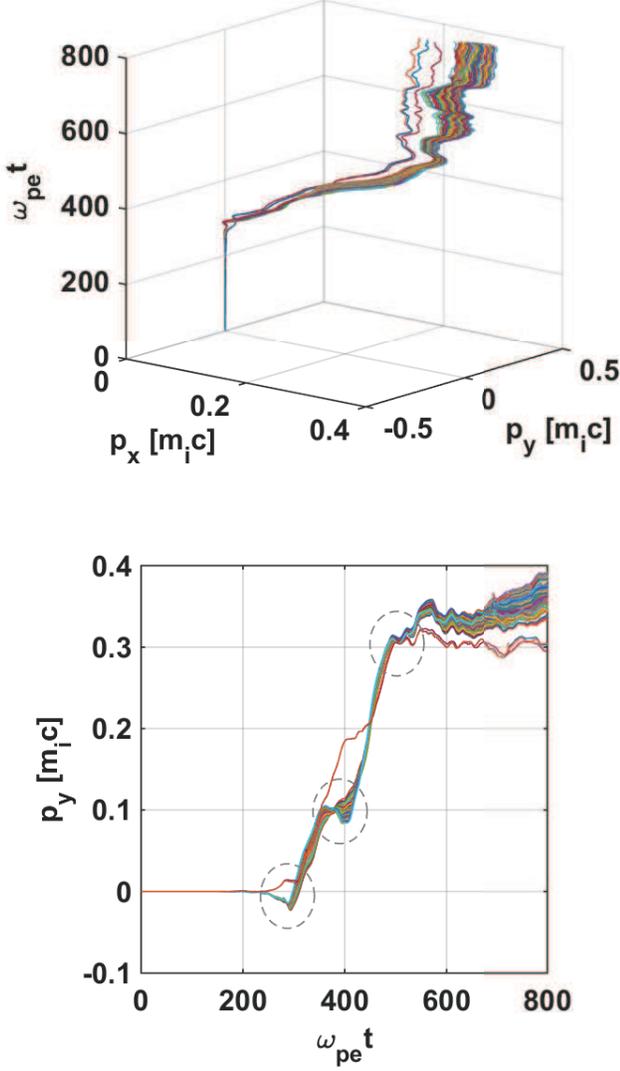}
\caption{The momentum space trajectory of 100 ions from background
plasma with the largest transverse momentum at
$t=950\omega_{pe}^{-1}$. } \label{fermi2trace}
\end{figure}

To interpret the strong transverse acceleration and the
inverse-power energy spectrum, we notice the strong
turbulence-generated magnetic fields. Fermi demonstrates that
particles can be accelerated by collisions with the moving
magnetic islands, and proves that for a large number of
collisions, the resulted energy spectrum will be in the form of
$dN/dE\sim E^{-[1+\tau/(B^2T)]}$ \cite{FermiIIOrigin}, where
$\tau$ is the average time between two collisions and $T$ is the
lifetime of the particle before being absorbed or escaped from the
region. This is often referred to as Fermi II acceleration.
Consider the following fact that (1) our code for this simulation
does not include the annihilate or recombination module, i.e., the
local particle number conserves, (2) a periodic boundary condition
is used in $y$-direction, and (3) the $x$-direction size of the
simulation domain is much larger for ions initially located at $x
< 400d_e$ to escape. It is thus reasonable to assume that
$T\rightarrow \infty$ and then $dN/dE\sim E^{-1}$, which is
exactly the result we get in the simulation. We also trace the
trajectories of the 100 ions in momentum space with the largest
transverse momentum at $t=950\omega_{pe}^{-1}$, as plotted in
Figure \ref{fermi2trace}. One can see obvious evidence of the
reflection between significant acceleration or deceleration marked
by the grey circles, which indicate that some ions experience
sequential acceleration occurs at different locations. This is
different from the usual acceleration mechanisms of ions found in
relativistic laser-solid interaction in laboratory \cite{SuLN}.
\par

We note that some researchers have reported some results from
similar models with much smaller incident flow energy
($\gamma_{flow} <10$) since one decade ago and have seen some
slight broadening of the energy spectra of the background plasma
particles. Because the maximum energy of the energetic particles
found in the simulations is small, there are controversial
explanations of this energy spectrum broadening. Some researchers
regard it as an acceleration\cite{Silva2003, Nishikawa2003}, while
others consider it as thermalization \cite{Frederiksen2004}. With
the continuous incidents of plasma flows with much higher initial
energy, our simulation enables the generation of stronger
turbulent magnetic fields and shows irrefragable evidence for the
existence of particle acceleration.

\subsection{Second stage: shock acceleration}
\begin{figure}
\includegraphics[width=0.48\textwidth]{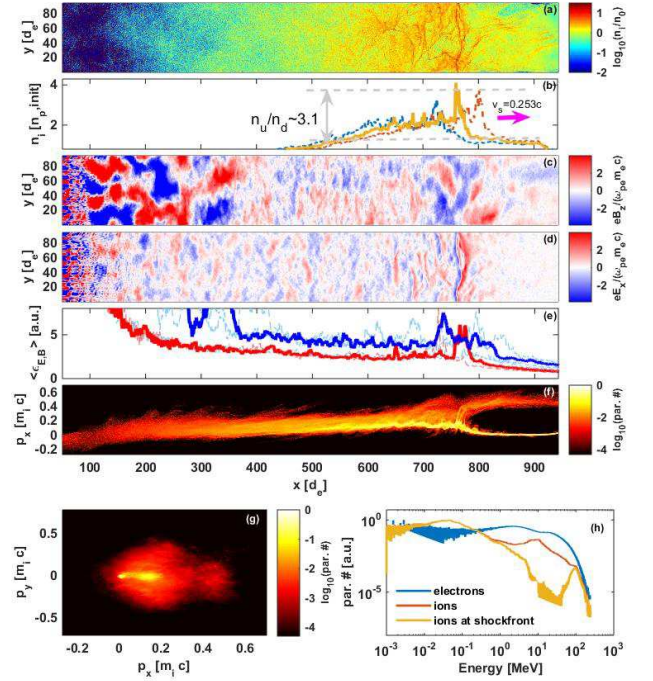}
\caption{(a) The space distribution of ion density, (b) the transverse averaged ion density, the space distribution of
(c) magnetic field $B_z$ and (d) electric field $E_x$, (e) transverse averaged field energy densities, blue for magnetic
field $B_z$ and red for electric field $E_x$, (f)$x-p_x$ and (g)$p_x-p_y$ projections of the ions' phasespace distribution
and the spectra of background particles at time $t = 2500\omega_{pe}^{-1}$. In panel (b) and (e), the dash lines shows the
corresponding quantity at $t=2470\omega_{pe}^{-1}$ and $t=2530\omega_{pe}^{-1}$ respectively (details are described in the text).} \label{shock}
\end{figure}

Apparently the lepton jet propagates much faster than the
ion-acoustic speed of the  background plasma, which is estimated
as around $v_a = \sqrt{(\gamma_{ad} T_{pe}/m_i)}\approx 0.13c$.
Here $\gamma_{ad}$ is the adiabatic coefficient which equals to 2
for a simple particle in a two-dimensional frame. Since the system
is highly disequilibrium, $T_{pe}$ is approximated with the
average energy of background electrons, which is about 7MeV
according to the simulation. As a result, a shock structure
gradually forms in the background plasma as the interaction
continues. Figure \ref{shock}(a) shows the ion density
distribution at $t = 2500\omega_{pe}^{-1}$ and its transverse
average is plotted in Figure \ref{shock}(b) with the cases of
30$\omega_{pe}^{-1}$ earlier and later (shown by the blue and
dark-red dash lines, respectively). One can see clearly that a
shock wavefront propagates forward at a speed of $v_S\simeq
0.253c$, which gives a Mach number $\mathscr{M} = 1.95$. And the
density ratio between downstream and upstream of the shock is
measured as around $n_{down}/n_{up} \sim 3.1$ in Figure
\ref{shock}(b). The shock hydrodynamic jump conditions give
\begin{gather}
n_{down}/n_{up} = (\gamma_{ad}\hat{\gamma}+1) / (\gamma_{ad}-1)\\
v_S = c
[(1+\gamma_{ad}\hat{\gamma})\hat{p}]/[1+\hat{\gamma}+\gamma_{ad}\hat{p}^2]
\end{gather}
for a well defined shock wave\cite{shocktheory, FiuzaPRL2012}, in
which $\gamma_{ad}$ is the adiabatic coefficient with value 2 as
discussed in the beginning of this paragraph. $\hat{p}$ denotes
the relative momentum (normalized by the rest mass multiplying
$c$) of the downstream flow in the frame of the upstream (which in
our case is the rest frame) and $\hat{\gamma}$ is the related
Lorentz factor. According to Figure \ref{shock}(f), we take
$\hat{p} \sim 0.17$ as a reasonable estimate, then $\hat{\gamma}
\sim 1.014$. Put these values into the above formulas, we obtain
$n_{down}/n_{up} = 3.03$ and $v_S = 0.248c$, which agree with the
measurement in our simulation.\par

Panels(c) and (d) of Figure \ref{shock} present the distributions
of electromagnetic fields. One can see that at $x < 400d_e$, the
turbulence-generated electromagnetic fields discussed in the
earlier part of the paper still remain strong although the ions
are almost blown up by the lepton flow. The field intensity
dramatically drops to some random fluctuations. However, at the
place of the shock wavefront, the fields become strong again and
form a peak. Interestingly, we find that the peak positions of the
electric and magnetic fields do not overlap with each other. The
former locates in front of the wavefront; the latter locates
behind the wavefront and exists in a broader range, which is
because the sources of the fields are different. The electric
field is generated by the charge separation in front of the wave
front while the source of the magnetic field is the instabilities.
The fields move with the shock and then reflect the ions at the
shock front, with  a speed of twice of the shock wave speed, as
one can confirm through the high branch in Figure \ref{shock}(f)
for the longitudinal phasespace. Figure \ref{shock}(g) shows the
$p_x-p_y$ projections of the ion's phasespace distribution. There
is no further transverse acceleration comparing with
$t=950\omega_{pe}^{-1}$, while the longitudinal acceleration is
significantly enhanced by the shock. We also plot the energy
spectra of the background electrons and ions in Figure
\ref{shock}(h). The spectrum of ions is not perfectly
inverse-power as a whole now, but there are piecewise
inverse-power parts at $10^{-1}-10^1$MeV and $10^{1}-10^2$MeV,
which forms a "knee" structure. Similar phenomena are found in
astronomical observations as well\cite{TerPRD2014, GaryakaAP2007}.
The orange line plots the spectrum of ions located at $x>800d_e$
(shockfront). It shows that a quasi-monoenergetic peak is formed
by the shock-reflected ions at about 200MeV, which is four times
larger than the energy of the injecting lepton flow. Here we would
like not to discuss the details of shock acceleration since it is
well studied in plenty of previous publications\cite{ShockS2,
AstroShock,ShockS3,YangJGR2009}.\par

\begin{figure}
\includegraphics[width=0.48\textwidth]{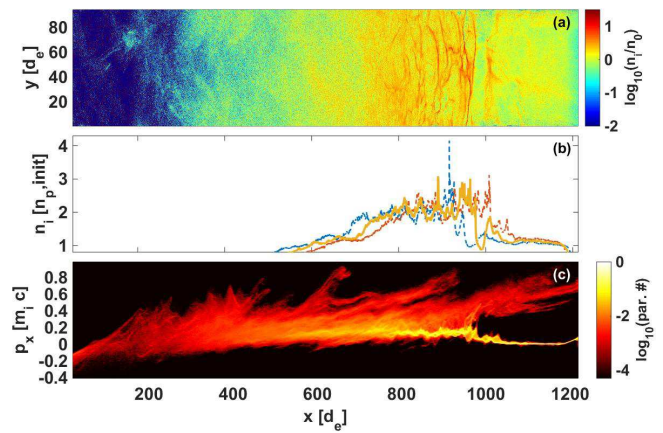}
\caption{\label{gamma200}(a) The space distribution of the
background ion density, (b) the transverse averaged ion density
and (c) $x-p_x$ projections of the ions' phasespace distribution
at time $t = 2500\omega_{pe}^{-1}$ for the initial 100MeV lepton
flow case.}
\end{figure}

To check the robustness of the above ion acceleration processes,
we have also performed another simulation with higher initial
energy of the  lepton flow of $\sim 100$MeV($\gamma_{e^+,e^-} =
200$) with other conditions unchanged. Similar phenomena are
observed. In the first stage, the protons, which are accelerated
by the Fermi II mechanism as well as momentum transfer from the
lepton flow longitudinally, form a perfect inverse-power spectrum.
The cut-off energy is slightly increased to $\sim 130$MeV.
However, this increase is caused only by the increase of
longitudinal momentum, which implies that the Fermi II mechanism
is not enhanced, i.e., the average speed of the "magnetic islands"
generated by the instabilities is insensitive to the initial
energy of the flow. On the other hand,  a lepton flow with higher
energy can provide more longitudinal momentum for the downstream
electrons, which significantly increases its temperature to
15.7MeV. As a result, the ion-acoustic speed increases to $0.18c$
and the measured shockwave speed increases to $v_S=0.293c$. The
ions reflected by this shock wave form a quasi-monoenergetic
structure at about 100-400MeV. However, for this
$\gamma_{e^+,e^-}=200$ case, we observe that some ions in the
downstream also have very high energy despite their small
population, as shown in Figure \ref{gamma200}(c), which is
slightly different from the $\gamma_{e^+,e^-}=100$ case. This may
lead to some injected acceleration into the shock from the
pre-accelerated ions in the downstream region. We also present the
space distribution of the ion density in Figures \ref{gamma200}(a)
and \ref{gamma200}(b) for comparison with the corresponding panels
in Figure \ref{shock}.\par

\section{Conclusion}
In conclusion, we have studied the interaction of a high-energy
lepton flow with a background normal  ion-electron plasma via PIC
simulation. A scenario of two-stage acceleration of the background
ions is identified: In the first stage, the main mechanism is
type-II Fermi acceleration found both in the transverse and
longitudinal directions. The accelerated ions forms a globally
inverse-power energy spectrum. As time passes, the acceleration
gradually enters the second stage as a shock wavefront forms in
the background plasma. The shock propagates in the longitudinal
direction with a Mach number of about 2, which further accelerates
the background ions to energy higher than that gained in the first
stage. While the ion spectrum is no longer globally inverse-power
in the second stage, it becomes piecewise inverse-power with a
knee structure. And the results, which will be relevant with the
interaction between cosmic flows with interstellar matters, are
helpful to explain the origin of the cosmic rays with
inverse-power spectrum. The acceleration scenario presented in
this work may be tested experimentally in the future as the high
energy lepton flows can be created nowadays by high power lasers
\cite{jet3,jet4,jet5}.

\vspace{2mm} This work was partially supported
by the National Basic Research Program of China (Grant No.
2013CBA01500), and the National Science Foundation of China (Grant
Nos. 11421064, 11220101002, 11129503, 11135012).



\begin{thebibliography}{99}

\bibitem{FermiIIOrigin}
Fermi E.
\newblock On the origin of the cosmic radiation.
\newblock {Phys. Rev.}, 1949, 75:1169--1174

\bibitem{InvCompton1}
{Aharonian F} et~al.
\newblock Discovery of the two ¡°wings¡± of the kookaburra complex in vhe
  $\gamma$-rays with hess.
\newblock {Astronomy \& Astrophys}, 2006, 456:245--251

\bibitem{InvCompton2}
{Aharonian F} et~al.
\newblock First detection of a vhe gamma-ray spectral maximum from a cosmic
  source: Hess discovery of the vela x nebula.
\newblock {Astronomy \& Astrophys}, 2006, 448:L43--L47

\bibitem{synchro}
Ginzburg V L and Syrovatsk S I.
\newblock Developments in the theory of synchrotron radiation and its
  reabsorption.
\newblock {Ann Rev Astrono \& Astrophys}, 1969, 7:375--420

\bibitem{BhattacharjeePR2000}
Bhattacharjee P and Sigl G.
\newblock Origin and propagation of extremely high-energy cosmic rays.
\newblock {Phys Rep}, 2000, 327:109--247

\bibitem{OlintoPR2000}
Olinto A V.
\newblock Ultra high energy cosmic rays: the theoretical challenge.
\newblock {Phys Rep}, 2000, 333¨C334:329--348

\bibitem{FermiIIRecon}
Hoshino M.
\newblock Stochastic particle acceleration in multiple magnetic islands during
  reconnection.
\newblock {Phys Rev Lett}, 2012, 108:135003

\bibitem{ShockSNR}
Blandford R and Eichler D.
\newblock Particle acceleration at astrophysical shocks: A theory of cosmic ray
  origin.
\newblock {Phys Rep}, 1987, 154:1--75

\bibitem{JofR}
Begelman M C, Rees M J, and Sikora M.
\newblock Energetic and radiative constraints on highly relativistic jets.
\newblock {Astrophys J}, 1994,429:L57--L60

\bibitem{WangB} Wang B, Yuan Q, Fan C,  et al.
\newblock A study on the sharp knee and fine structures of cosmic ray spectra
\newblock {Sci China -- Phys Mech Astron}, 2010, 53:842--847

\bibitem{GRB1}
Zhang B and Meszaros P.
\newblock Gamma-ray bursts: Progress, problems \& prospects.
\newblock {Int J Mod Phys A}, 2004, 19:2385--2472

\bibitem{GRB2} Piran T.
\newblock The physics of gamma-ray bursts.
\newblock {Rev Mod Phys}, 2005,76:1143--1210

\bibitem{FofUni1}
Loeb A and Waxman E.
\newblock Cosmic [gamma]-ray background from structure formation in the
  intergalactic medium.
\newblock {Nature}, 2000, 405:156--158

\bibitem{FofUni2}
Gruzinov A.
\newblock Gamma-ray burst phenomenology, shock dynamo, and the first magnetic
  fields.
\newblock {Astrophys J Lett}, 2001, 563:L15--L18

\bibitem{ShockA1}
Ucer D and Shapiro V D.
\newblock Unlimited relativistic shock surfing acceleration.
\newblock {Phys Rev Lett}, 2001, 87:075001

\bibitem{AstroShock}
Haugb{\o}lle T.
\newblock Three-dimensional modeling of relativistic collisionless ion-electron
  shocks.
\newblock {Astrophys J Lett}, 2011, 739(2):L42-L45.

\bibitem{ShockS2}
Martins S F, Fonseca R A, Silva L O, and Mori W B.
\newblock Ion dynamics and acceleration in relativistic shocks.
\newblock {Astrophys J Lett}, 2009, 695:L189--L192

\bibitem{ShockS3}
Amato E and Arons J.
\newblock Heating and nonthermal particle acceleration in relativistic,
  transverse magnetosonic shock waves in proton-electron-positron plasmas.
\newblock  {Astrophys J}, 2006, 653:325--338

\bibitem{YangJGR2009}
Yang Z Y, Lu Q M, Lemb{\`e}ge B, and Wang S.
\newblock Shock front nonstationarity and ion acceleration in supercritical
  perpendicular shocks.
\newblock {J Geophys Res: Space Physics}, 2009,
  114(A3):2156--2202

\bibitem{pulsarwind}
{van der Swaluw E}, {Achterberg A}, {Gallant Y A}, {Downes T P},
and {Keppens R}.
\newblock Interaction of high-velocity pulsars with supernova remnant shells.
\newblock {Astronomy \& Astrophys}, 2003, 397:913--920


\bibitem{Fireball} Waxman E.
\newblock Gamma-ray bursts and collisionless shocks.
\newblock {Plasma Phys Control Fusion}, 2006, 48:B137


\bibitem{BinaryOrbit}
Chomiuk L, Linford J D, Yang J, et al.
\newblock Binary orbits as the driver of [ggr]-ray emission and mass ejection
  in classical novae.
\newblock {Nature}, 2014, 514:339--342


\bibitem{jet1} Wardle J F C, Homan D C, Ojha R and Roberts D H,
\newblock Electron¨Cpositron jets associated with the quasar 3C279,
\newblock {Nature},  1998, 395: 457--461
\bibitem{jet2} Ruffini R, Vereshchagin G, Xue S S,
\newblock Electron¨Cpositron pairs in physics and astrophysics: From heavy nuclei to black holes,
\newblock {Phys Rep},  2010, 487: 1--140
\bibitem{CuiYQ2015} Cui Y Q,
\newblock Numerical studies on selected problems in
the interaction of intense lasers/particle beams with bulk plasma,
\newblock {PhD Thesis, University of Chinese Academy of Sciences}, May 2015

\bibitem{WangPRE2015}
Wang W M, Gibbon P, Sheng Z M, and Li Y T.
\newblock Integrated simulation approach for laser-driven fast ignition.
\newblock {Phys Rev E}, 2015, 91:013101

\bibitem{Mondalturb}
Mondal S, Narayanan V, Ding W J, Lad A D, Hao B, Ahmad S,
  Wang W M, Sheng Z M, Sengupta S, Kaw P, Das A, and
  Kumar G R.
\newblock Direct observation of turbulent magnetic fields in hot, dense laser
  produced plasmas.
\newblock {Proc Nat Acad Sci}, 2012, 109:8011--8015

\bibitem{SuLN} Su L N, et al.
\newblock Proton angular distribution research by a new angle-resolved proton energy spectrometer
\newblock {Sci China -- Phys Mech Astron}, 2014, 57:844--848

\bibitem{Silva2003}
Silva L O, Fonseca R A, Tonge J W, et al.
\newblock Interpenetrating plasma shells: Near-equipartition magnetic field
  generation and nonthermal particle acceleration.
\newblock {Astrophys J}, 2003, 596:L121-L124

\bibitem{Nishikawa2003} Nishikawa K I, Hardee P, Richardson G, Preece R, Sol H, and
  Fishman G J.
\newblock Particle acceleration in relativistic jets due to weibel instability.
\newblock {Astrophys J}, 2003, 595:555--563

\bibitem{Frederiksen2004}
Frederiksen J T, Hededal C B, Haugb\o~lle T, and Nordlund A.
\newblock Magnetic field generation in collisionless shocks: Pattern growth and
  transport.
\newblock {Astrophys J}, 2004, 608:L13

\bibitem{shocktheory}
Blandford R D and McKee C F.
\newblock Fluid dynamics of relativistic blast waves.
\newblock {Phys Fluids}, 1976, 19:1130--1138


\bibitem{FiuzaPRL2012}
Fiuza F, Fonseca R A, Tonge J, Mori W B, and Silva L O.
\newblock Weibel-instability-mediated collisionless shocks in the laboratory
  with ultraintense lasers.
\newblock {Phys Rev Lett}, 2012, 108:235004


\bibitem{TerPRD2014}
Ter-Antonyan S.
\newblock Sharp knee phenomenon of primary cosmic ray energy spectrum.
\newblock {Phys Rev D}, 2014, 89:123003

\bibitem{GaryakaAP2007}
Garyaka A P, Martirosov R M, Ter-Antonyan S V, Nikolskaya N,
Gallant Y A, Jones L, and Procureur J.
\newblock Rigidity-dependent cosmic ray energy spectra in the knee region
  obtained with the gamma experiment.
\newblock {Astroparticle Phys}, 2007, 28:169--181


\bibitem{jet3} Gahn C, Tsakiris G D, Pretzler G, et al.
\newblock  Generating positrons with femtosecond-laser pulses,
\newblock {Appl Phys Lett}, 2000, 77: 2662--2664

\bibitem{jet4} Chen H, Wilks S C, Meyerhofer D D, et al.
\newblock  Relativistic quasimonoenergetic positron jets from intense laser-solid interactions,
\newblock {Phys Rev Lett}, 2010, 105: 015003

\bibitem{jet5} Sarri G, Poder K, Cole J M, et al.
\newblock  Generation of neutral and high-density electron¨Cpositron pair plasmas in the laboratory
\newblock {Nature Comm}, 2015,  6: 6747




\end{thebibliography}
\end{document}